\documentclass[aps,pre,showpacs,showkeys,twocolumn,superscriptaddress,
floats,floatfix,preprintnumbers,longbibliography]{revtex4-1}
\usepackage{amssymb,amsmath}
\usepackage{bm,url,graphics,subfigure,epsfig,rotating,float}
\usepackage{soul}
\usepackage{mathrsfs}
\usepackage{hyperref}
\usepackage[normalem]{ulem}   % for striking out
\usepackage[dvipsnames, table, xcdraw]{xcolor}

\hypersetup{ pdfauthor={VA VP DM}, pdfsubject={Physics},
            colorlinks=true, linkcolor=Maroon, citecolor=Maroon,
        urlcolor=blue }

\usepackage{color,soul}
\graphicspath{{./}{fig/}}
\usepackage{relsize}
\def \VLJ   {V_{\rm LJ}}
\def \VRLJ  {V_{\rm RLJ}}
\def \sigA  {\sigma_{\rm A}}
\def \epsA  {\epsilon_{\rm A}}
\def \sigw  {\sigma_{\rm w}}
\def \epsw  {\epsilon_{\rm w}}
\def \mhat {\hat{\bm m}}
\def \RRij {{\bm R}_{\rm ij}}
\def \RRijm {{\bm R}_{\rm ij-1}}
\def \Es    {E_{\rm s}}
\def \Eb    {E_{\rm bend}}
\def \Ebl   {E_{\rm bulk}}

\def \ET    {\mbox{ET}}

\def \Yt    {Y_{\rm 3d}}
\def \sigmat {\sigma_{\rm 3d}}

\def \deli  {\partial_{i}}
\def \delj  {\partial_{j}}

\def \mE {\mathcal{E}}

\def \aa {a}

\def \uu  {{\bm u}}

\def \lzij  {\ell^{\rm 0}_{\rm ij}}
\def  \Rij  {R_{\rm ij}}
\def  \Rijm  {R_{\rm ij-1}}
\def  \Ri  {\bm{R}_{\rm i}}
\def  \Rj  {\bm{R}_{\rm j}}

\def  \RR  {{\bm R}}

\def \lap {\nabla^2}

\def \Fv  {\mbox{FvK}}

\def \kB {k_{\rm B}}

\def \ie  {i\/.e\/.}

\newcommand{\eq}[1]{~(\ref{#1})}

\newcommand{\fig}[1]{Fig.~(\ref{#1})}

\newcommand{\subfig}[2]{Fig.~(\ref{#1}#2)}

\newcommand{\bfig}{\begin{figure}}
\newcommand{\efig}{\end{figure}}
\newcommand{\bc}{\begin{center}}
\newcommand{\ec}{\end{center}}
\newcommand{\bea}{\begin{eqnarray}}
\newcommand{\eea}{\end{eqnarray}}

\def \rB {{\rm{bend}}}
\def \cA {{\mathcal {A}}}

\def \nhat {\hat{{\bm n}}}

\def \cAib {{\mathcal A}_i}
\def \aij {\alpha_{\rm ij}}
\def \bij {\beta_{\rm ij}}
\def \bijm {\beta_{ij-1}}
\def \zero {{\rm 0}}

\def \LL {{\bm L}}
\def \Sum {\mathlarger{\sum}}

\def \Kl {K_{\rm L}}
\begin{document} 
\title{MeMC: A package for monte-carlo simulations of
spherical shells}

\author{Vipin Agrawal}
\affiliation{ Nordita, KTH Royal Institute of Technology and
  Stockholm University, Roslagstullsbacken 23, 10691 Stockholm, Sweden}
\affiliation{ Department of Physics, Stockholm university, Stockholm,Sweden.}
\author{Vikash Pandey}
\affiliation{ Nordita, KTH Royal Institute of Technology and
Stockholm University, Roslagstullsbacken 23, 10691 Stockholm, Sweden}
\author{Hanna Kylhammar}
\affiliation{ KTH Royal Institute of Technology, Sweden}
\author{Apurba Dev}
\affiliation{ Department of Electrical Engineering, The Angstr\"om Laboratory, 
Uppsala University, Uppsala, Sweden.}
\affiliation{Department of Applied Physics, School of Engineering Sciences, 
KTH Royal Institute of Technology, Stockholm, Sweden.}
\author{Dhrubaditya Mitra}
\email{dhruba.mitra@gmail.com}
\affiliation{ Nordita, KTH Royal Institute of Technology and
Stockholm University, Roslagstullsbacken 23, 10691 Stockholm, Sweden}

%The abstract will become the summary in the JOSS paper
\begin{abstract}
  The MeMC is an open-source software package for monte-carlo simulation of
  elastic shells. It is designed as a  tool to interpret the force-distance
  data generated by indentation of biological nano-vesicles by atomic force
  microscopes.  The code is written in c++ and python. The code is
  customizable -- new modules can be added in a straightforward manner.
\end{abstract}
\maketitle
%
%\section{Summary}

\section{Statement of need and purpose of software}
Micro and nano vesicles, both natural and synthetic, play a crucial
role in biology and medicine.
The physical properties of these vesicles play an important role in
their biological functions~\cite{phillips2012physical}.
Hence it is important to be able to
measure their elastic constants, in particular the Young's modulus and
the bending rigidity. One way to measure the elastic constants of biological
objects, e.g., a red blood cell (RBC), is to poke them with an atomic force
microscope (AFM) to obtain a force-distance curve.
Then we must model the biological object as an elastic material and
by fitting this model to the experimental force-distance curve to estimate
the parameters of the elastic model, i.e., the elastic constants.
As an example, consider a force-distance curve obtained by
AFM measurements of an RBC. The RBC is modeled as a linear elastic material
with a Young's modulus, $\Yt$. Typically a Hertzian model of elastic bodies
in contact~\cite[][section 9]{LLelast} is used to measure $\Yt$.
Nano vesicles differ from (micro-meter scale) cells in two important
ways
\begin{enumerate}
\item The nano-vesicles are much smaller hence thermal fluctuations
  may effectively renormalize the elastic
  coefficients~\cite{paulose2012fluctuating,kovsmrlj2017statistical}. 
\item Cell membranes are strongly coupled to an underlying cytoskelton.
  Hence they may be modeled by a solid body~\cite{hw2002stomatocyte} but
  nano-vesicles must be modeled as liquid filled elastic membranes. 
\end{enumerate}
Hence, to be able to interpret the force-distance curve of nano-vesicles,
we need to solve for the elastic response of a thermally fluctuating
elastic shell. 

There are commercial packages, e.g., COMSOL~\cite{comsol}, 
to calculate the force-distance
curve of solid bodies and closed membranes with fluids inside
under the action of external forces
but to the best of our knowledge there is no package that  includes thermal
effects, which are important in nano-vesicles. 
Monte-carlo simulations of elastic membranes, that does include
thermal fluctuations, have been done for
more than three
decades~\cite{goetz1999mobility, bowick2001universality,
  auth2005fluctuation, paulose2012fluctuating},
see also \cite[][for a review ]{gompper2004triangulated}.
But to the best of our knowledge there is no open-source code available.
The goal of this package is to fill this gap in open-source software. 
\section{Theoretical background}

Our model of nano-vesicles is an homogeneous amorphous  membrane enclosing
an incompressible fluid~\cite{vorselen2017competition}.
Unlike a solid ball, the force-distance relationship for such a model is
linear for small deformation~\cite{vorselen2017competition,paulose2012fluctuating}
if we ignore thermal fluctuations. 
Ref.~\cite{vorselen2017competition} uses a similar model, ignoring
thermal fluctuations,
to interpret AFM measurement of nano-vesicles.

Let us consider a (three dimensional) material with Young's
modulus $\Yt$ and Poisson's ratio $\sigmat$ and make a membrane out of it.
Then the bending modulus and the in-plane Young's modulus
are ~\cite[][section 13 and 14]{LLelast}
\begin{equation}
  B = \frac{\Yt h^3}{12(1-\sigmat^2)}\quad\text{and}\quad Y = \Yt h\/,
  \label{eq:2d3d}
\end{equation}
where $h$ is the thickness of the membrane.
This need not necessarily hold for biological membranes.
Nevertheless consider a fluid enclosed by a solid membrane, as done in
\cite{paulose2012fluctuating}. We  consider an elastic energy of the form
\begin{subequations}
\begin{align}
  \mE\left[w,\uu\right] &= \int d^2x\left[ \frac{B}{2}\left( \lap w \right)^2
             + \mu u_{ij}^2 + \frac{\lambda}{2}u_{kk}^2 - pw \right] \\ 
  u_{ij} &= \frac{1}{2}\left( \deli u_j + \delj u_i + \deli w\delj w\right)
            -\delta_{ij}\frac{w}{R}
\end{align}
\end{subequations}
where $w$ is the out-of-plane deformation of the shell, and $\uu$ is the
in-plane deformation, $p$ is the pressure,
$\lambda$ and $\mu$ are the two in-plane L\'ame coefficients
and $B$ is the bending modulus.
The L\'ame coefficients are related to other elastic constant in the following manner~\cite{LLelast}
\begin{subequations}
  \begin{align}
    K &= \lambda + \frac{2}{3}\mu \\
    Y &= \frac{9K\mu}{3K+\mu} \\
    \sigma & = \frac{1}{2}\frac{3K-2\mu}{3K+\mu}
  \end{align}
\end{subequations}
Here $K$ is the volume compressibility, $Y$ the Young's modulus, and
$\sigma$ the Poisson ratio.

If we consider the material to be incompressible, $K \to \infty$
and $\sigma = 1/2$, then there are two
elastic constant, the bending rigidity $B$ and the Young's modulus $Y$.
Consequently, there are two dimensionless numbers, the  F\"oppl--von-Karman number
\begin{equation}
  \Fv = \frac{YR^2}{B}
  \label{eq:fvk}
\end{equation}
and the Elasto-Thermal number:
\begin{equation}
    \ET = \frac{\kB T}{B}\sqrt{\Fv}
\end{equation}
where $R$ is the radius of the spherical shell.

Using values of $Y$ and $B$ from molecular dynamics simulations
of lipid bilayers~\cite{boek2005mechanical}, $Y = 1.7$N/m and
$B = 5 \kB T$ and $R = 100$ nano meter,
we have $\Fv \approx 0.3\times 10^7 $ which is close to the 
F\"oppl--von-Karman number for Graphene sheets
and $\ET \approx 10^3$. 

\section{Numerical implementation}

\subsection{Grid}
\label{subsect:grid}
\begin{figure}
\includegraphics[width=0.65\columnwidth]{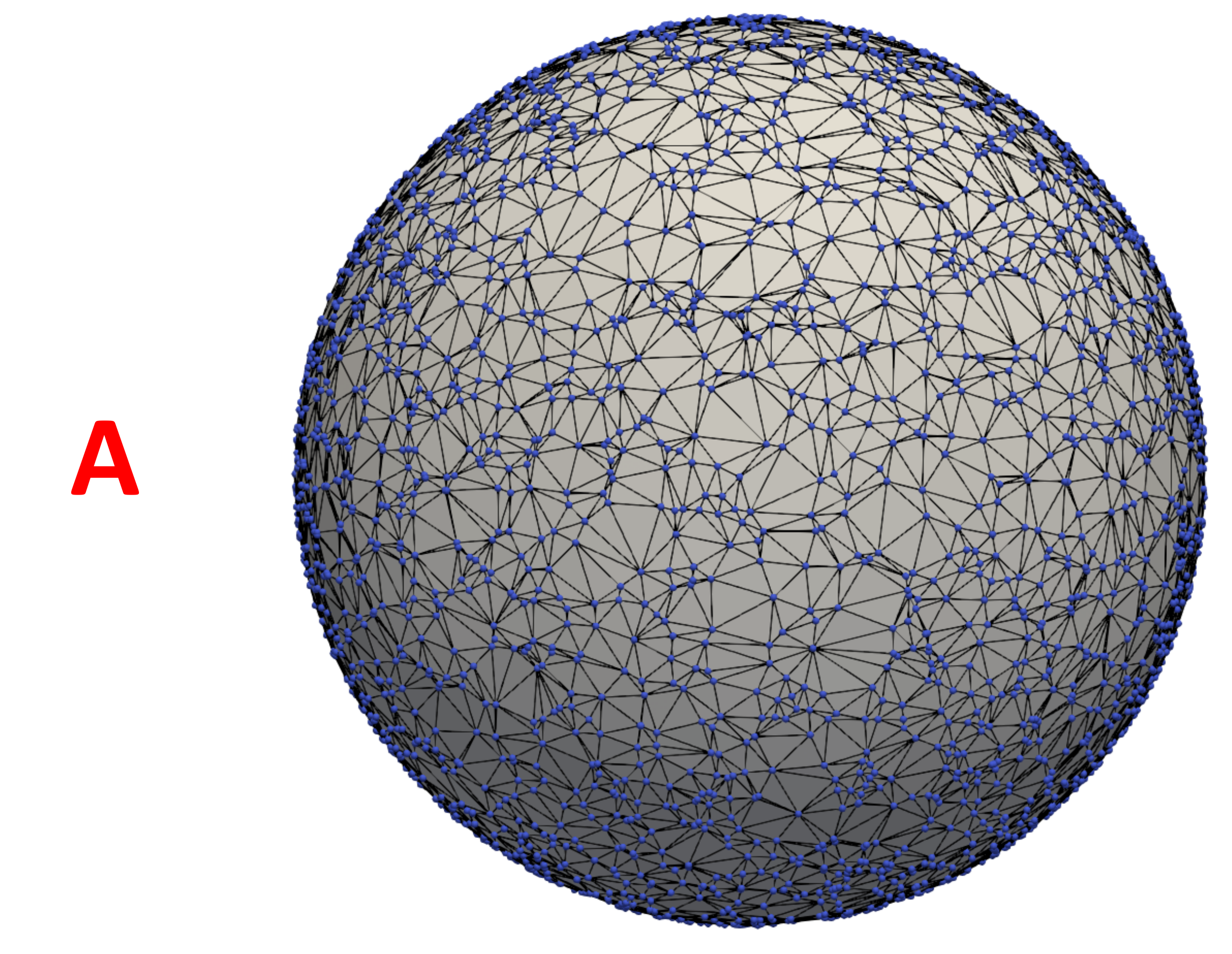}\\
\includegraphics[width=0.65\columnwidth]{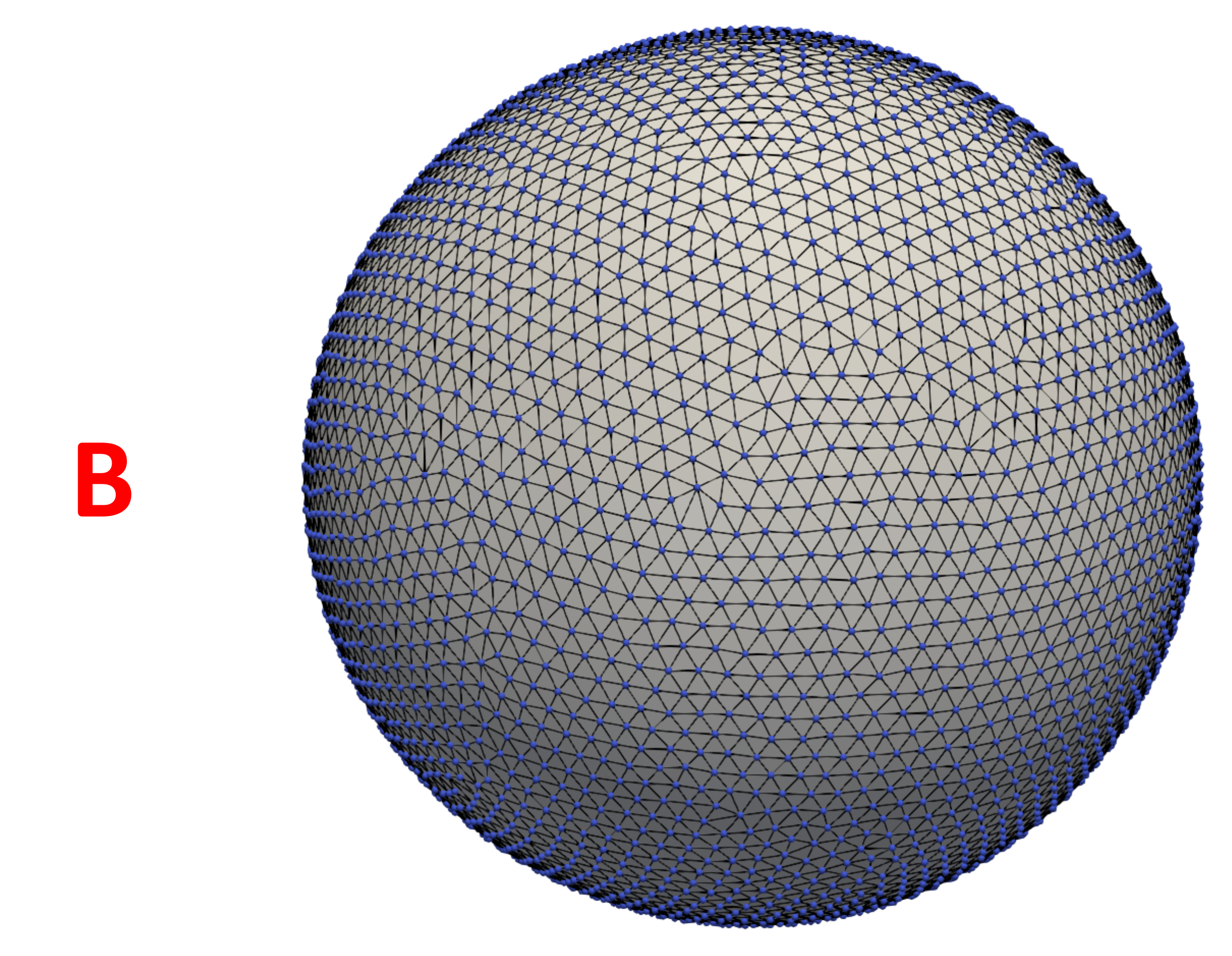} 
\includegraphics[width=0.65\columnwidth]{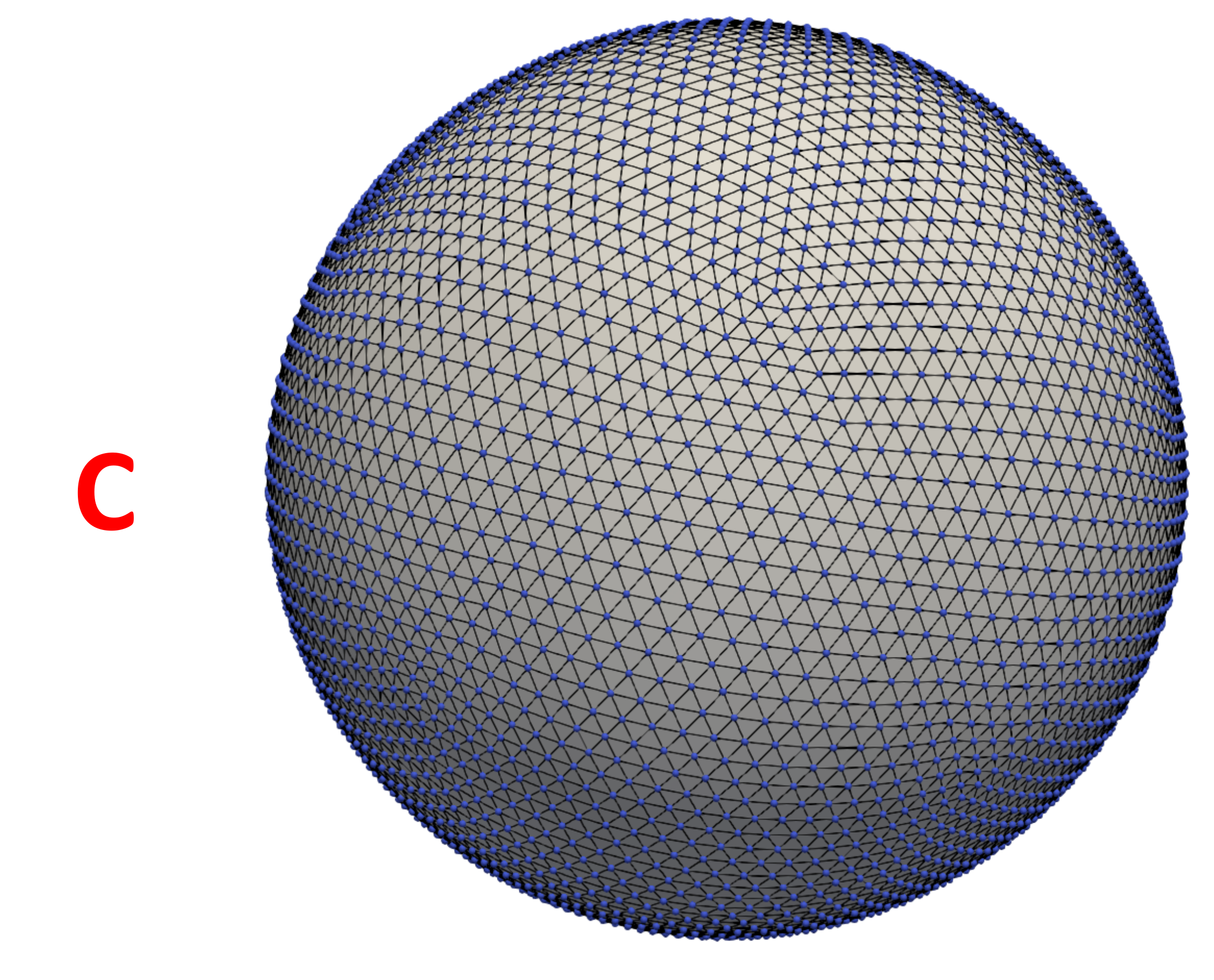} 
\caption{{\bf Grid points on a sphere} (A)Triangulated random points on a sphere.
  (B) Triangulated points on sphere after $60000$ SMC iteration of
  the initial configuration shown in (A). 
  (C) An example of regular grid }
\label{fig:grid}
\end{figure}

Following Ref.~\cite{gompper2004triangulated}, we use a triangulated-network
grid in the following manner. We start with $N$ randomly chosen points on a
sphere~\subfig{fig:grid}{A}.
Then, we run a Monte-Carlo simulation, with a Lennard-Jones (LJ) repelling
potential, of these points moving on the surface.
Once the surface Monte-Carlo (SMC)
has reached an equilibrium,
we use the algorithm in Ref.~\cite{caroli2010robust}
to construct the Delaunay triangulation of these points.
The connection between the points thus formed is kept unchanged.
In the rest of this paper we call this the \textit{initial configuration}. 
A different snapshot from the same equilibrium gives an equivalent but
differently triangulated grid.

An alternative is to use a regular grid~\cite{vliegenthart2006mechanical,buenemann2008elastic}. This is achieved by approximating
the sphere with geodesic polyhedron [\subfig{fig:grid}C] .
They can be generated using \textit{Meshzoo}
library \cite{meshzoo}.
In this paper we use $N=5120$ for the random grid and $N=5292$ for the
regular grid. 
%----------------------
%------------------------
\subsection{Energy}
The basic algorithm of the Monte-Carlo simulations is straightforward and
well-known~\cite[see, e.g.,][]{baumgartner2013applications}.
We randomly choose a point on the grid and move it by a random amount.
We calculate the change in energy due to this movement.
We accept or reject the move by the standard Metropolis algorithm.
In our code the  energy has the following contributions
\begin{equation}
   E = \Es + \Eb + \Ebl
  \label{eq:dE}
\end{equation}
where $\Es$ is the contribution from stretching, $\Eb$ is the
contribution from bending, and $\Ebl$ is the contribution from the bulk modulus.
Below, we describe each one of them in turn.
% We describe each of these in turn.
\subsubsection{Stretching}
In the initial configuration, two neighboring points with coordinates $\Ri$ and
$\Rj$ are connected by a bond of length $\lzij$. When the
$i$-th point is moved, all its bonds with the neighbors change from their
initial length. We model each of these bonds by a harmonic spring
and calculate the stretching energy by
\begin{subequations}
  \begin{align}
	  \Es &= \frac{1}{2}\Sum_i \frac{H}{2}\Sum_{j(i)}  \left(\Rij - \lzij\right)^2 \quad\text{where}
    \label{eq:Es} \\
    \Rij &\equiv \lvert \Ri - \Rj\rvert \/. 
    \label{eq:deltaEs}
  \end{align}
\end{subequations}
Here the notation $j(i)$ denotes that the sum is over all the nearest
neighbors of the $i$-th point. 

The L\'ame coefficients ($\lambda,\mu$) and the Young's modulus ($Y$) are given by~\cite{seung1988defects}
\begin{equation}
\lambda=\mu=\frac{\sqrt{3}}{4}H\/,\quad Y = \frac{2}{\sqrt{3}}H .
  \label{eq:klame}
\end{equation}
\subsubsection{Bending}
%---------------------------
\begin{figure}
  \includegraphics[width=0.8\columnwidth]{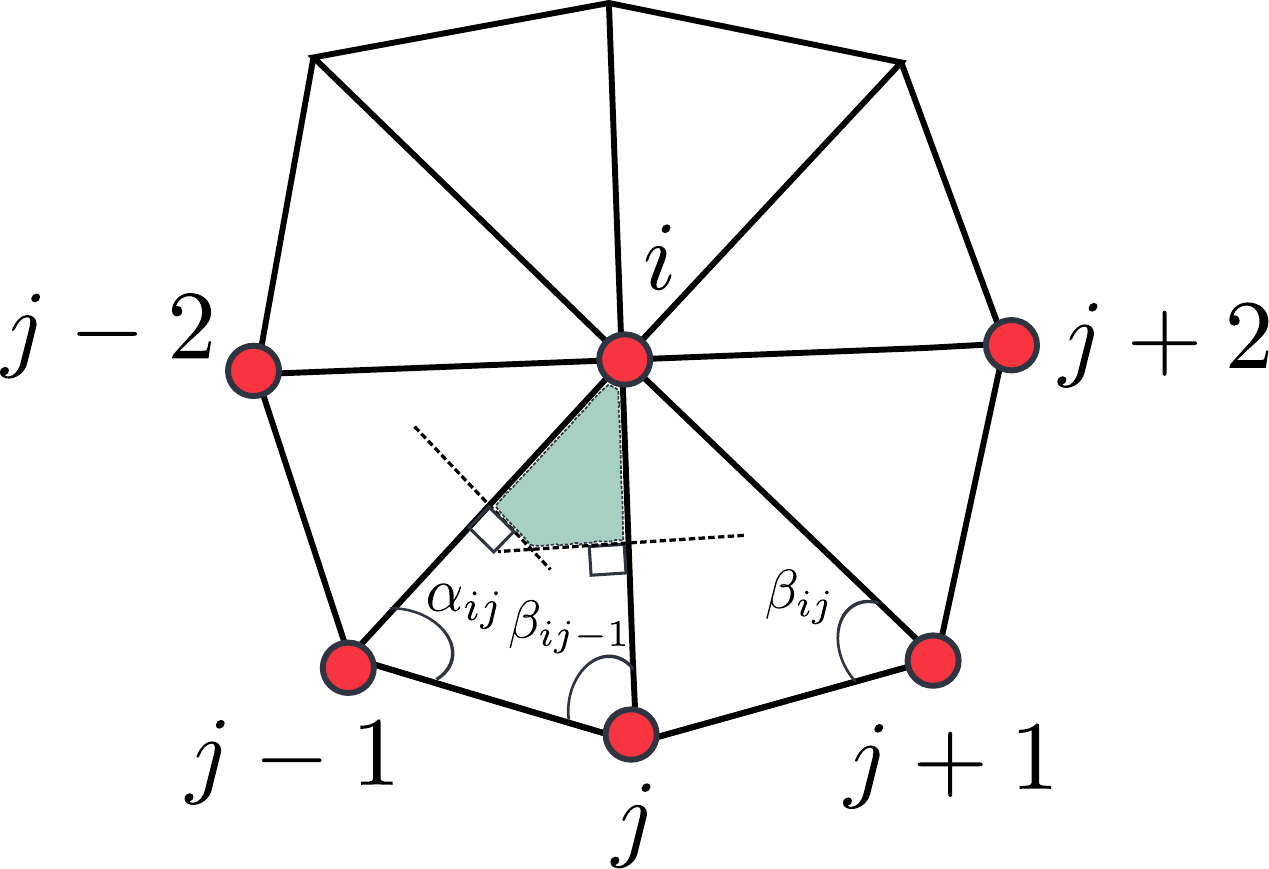}
  \caption{An example of triangulated mesh at the node $i$. 
  $\aij,\bij$ are the angles opposite to the bond $ij$.
    Shaded part is the  Voronoi region of triangle $T$ which is defined by
    the nodes ($i,j-1,j$).
  Here, we consider that the triangle $T$ is acute.}
  \label{fig:Voronoi}
\end{figure}
%---------------------------
To calculate the bending contribution,
we need to calculate the curvature.
In the continuum limit, \ie\ $N\to\infty$,
bending energy \cite{nelson2004statistical} is:
\begin{equation}
  E_{\rB} = \frac{B}{2}\int (\lap \RR)^2 dS,
  \label{eq:bendE_conti}
\end{equation}
where $\RR$ is the vector of a point on the surface of the sphere, $\lap \RR$ is Laplacian of $\RR$ on the surface of the sphere.

A general introduction to discretization of Laplacian on a triangulated mesh is
given in Refs.\cite{itzykson1986proceedings,hege2003visualization}.
Laplacian on a $2$D manifold embedded in $\mathbb{R}^3$ is:
\begin{equation}
  \LL(\RR)= 2\kappa(\RR)\mhat(\RR),
  \label{eq:laplace_beltrami}
\end{equation}
where $\kappa(\RR)$ is the mean curvature, and $\mhat(\RR)$ is the normal to
the surface at $\RR$.
Note that, $\mhat(\RR)$ is a local property of a point P with
coordinates $\RR$ and
it is not necessarily the outward normal of the
closed surface.
In the discrete form \cite{hege2003visualization,meyer2003discrete},
\begin{equation}
  \LL_i = \frac{1}{\cAib}\mathlarger{\mathlarger{\sum_{j(i)}}}
  \frac{1}{2}\left[\cot(\aij) + \cot(\bij)\right]\RRij\/.
\end{equation}
Here, $\cAib$ is the area of Voronoi dual cell at the node $i$, and
$\aij,\bij$ are the angles opposite to bond $ij$ as shown in \fig{fig:Voronoi}.
Consider the triangle T in \fig{fig:Voronoi} defined by its vertices
$(i,j-1,j)$.
If T is non-obtuse its circumcenter lies within it,
hence so does the Voronoi region.
Let $\cA^{\rm c}$ be the area of shaded region
  in \fig{fig:Voronoi}
  given by~\cite{meyer2003discrete,hege2003visualization},
\begin{equation}
  \cA^{\rm{c}} = \frac{1}{8} \left[\Rij^2\cot(\aij) + \Rijm^2\cot(\bijm) \right]\/.
  \label{eq:cA}
\end{equation}
If there is an obtuse angle in triangle T, the Voronoi region is not
necessarily enclosed by the triangle~\cite{hege2003visualization}.
For such cases, instead of $\cA^{\rm{c}}$, we use $\cA^{\rm{b}}$, defined
as~\cite{hege2003visualization,meyer2003discrete}:
\begin{equation}
  \cA^{\rm{b}} =
  \left\{
  \begin{array}{ll}
    \frac{\rm{area}(T)}{2}, \rm{angle\ of\ T\ at\ } i \rm{\ is\ obtuse} \\
    \frac{\rm{area}(T)}{4},  \rm{any\ other\ angle\ is\ obtuse} 
  \end{array}
\right\},
  \label{eq:bary_area}
\end{equation}
where ${\rm area}(T) = 0.5\left\lvert\RRij \times \RRijm\right\lvert$ is the
area of the triangle T.
The area $\cAib$ is obtained by summing up the contributions from
all the triangles in \fig{fig:Voronoi}, e.g., the contribution from
the triangle T is the shaded area. 

For a closed surface, the bending energy must be
calculated relative to the spontaneous curvature, i.e., its
discretised form is 
\begin{subequations}
  \begin{align}
    E_{\rB} &= \frac{B}{2} \mathlarger{\mathlarger{\sum_i}}\cAib
    \left(\LL_i - C\nhat\right)^2.
    \label{eq:bendE}
  \end{align}
\end{subequations}
where $C$ is the spontaneous curvature,
% Here $L^\zero$ is the spontaneous curvature of the surface,
for a sphere $C = 2/R$,
where $R$ is radius of the sphere and $\nhat$ is the outward normal to
the surface.
Hence not only the magnitude but also the vector nature of the
surface Laplacian must be determined.
For every triangle in the initial configuration, i.e., when
all the points lie on the surface of a sphere, the outward unit normal
can be calculated in a straightforward manner.
For example, for the triangle T in \fig{fig:Voronoi} it can be calculated
by finding out the unit vector that points along $\RRijm\times\RRij$.
Hence, at any time, if we access the points around the node i
in counterclockwise manner when viewed from outside we are guaranteed to
obtain the outward normal.
We ensure this by sorting appropriately the points around every
node in the initial configuration. As the connectivity of the mesh
remains unchanged this property is preserved at all future times.

To sort the neighbors around any node $i$, we rotate the coordinate system
such that, the $z$ axis passes through the point i along the vector $\RR_i$.
In this coordinate system we sort the neighbors by their
azimuthal angle.

Note that unlike Ref.~\cite{gompper2004triangulated}  we do not
incorporate self-avoidance. 
\subsubsection{Bulk}
We assume that the liquid inside the vesicle is incompressible ~\footnote{This
  is different from assuming a semi-permeable membrane,
  as done in Ref.~\cite{vorselen2017competition}, in which case
the liquid can flow in or out and the osmotic pressure of solutes
decreases and increases accordingly.}.
This is implemented by adding a energy cost to the volume change.
At any point, the total contribution to the bulk energy is
\begin{equation}
  \Ebl = \Kl\left(\frac{V}{V_\zero} - 1\right)^2\/,
\end{equation}
where $\Kl$ is bulk modulus, $V$ is current volume, and $V_\zero$ is the
undeformed initial volume of the vesicle.
As we move the point $i$ by a random amount, the
change in bulk energy is 
\begin{equation}
  \Delta \Ebl = 2\Kl\left(\frac{\Delta V(V  - V_0)}{V_0^2}+ \left(\frac{\Delta V}{V_0}\right)^2\right),
\end{equation}
where $\Delta V$ is the change in volume due to the move. 
\subsubsection{Pressure}
With addition of contribution from pressure difference from inside and
outside the shell our code can also be used for pressurized shells.
\subsection{Sticking to a solid surface}
\begin{figure}
  \includegraphics[width=0.4\columnwidth]{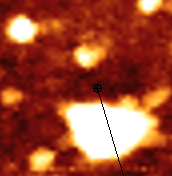}
  \put(-60,108){\textbf{\large (A)}}
  \qquad\qquad
  \includegraphics[width=0.25\columnwidth]{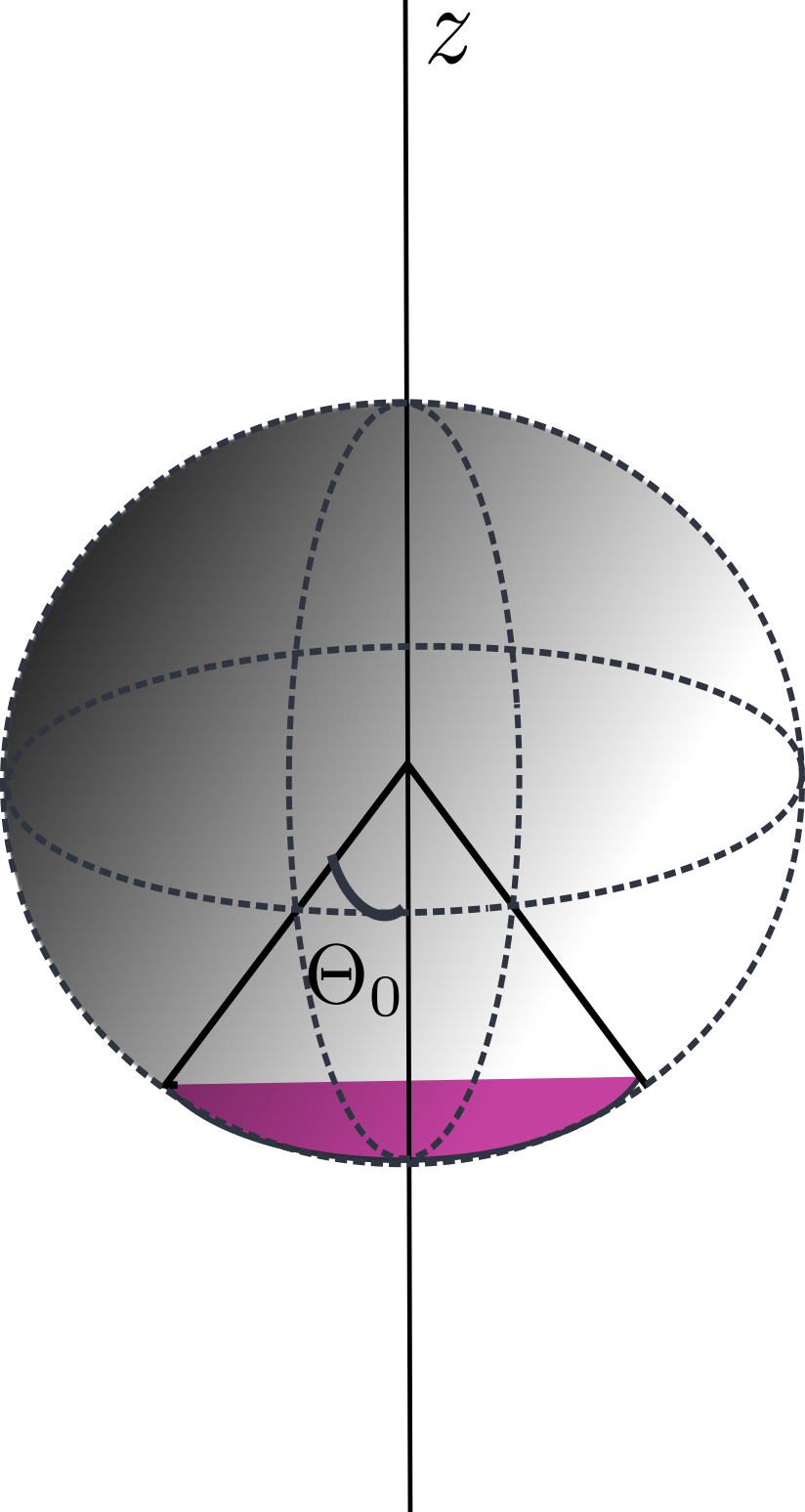}
  \put(-60,90){\textbf{\large (B)}}
  \\
  \includegraphics[width=\columnwidth]{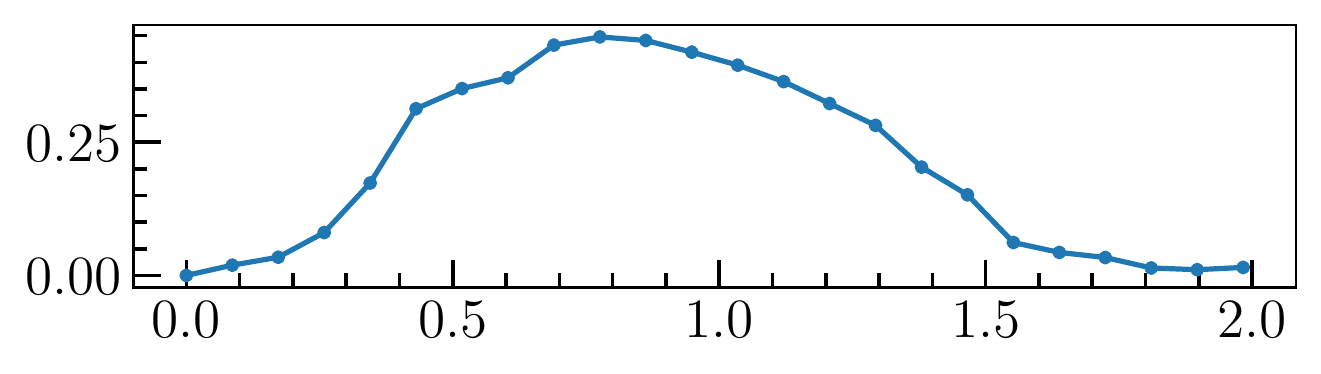}
  \put(-210,50){\textbf{\large (C)}}
  \caption{\label{fig:exo} (A) A colormap of the height as measured by the AFM.
  (B) Illustration of the vesicle stuck to a flat surface by an angle $\Theta_0$.
    (C) The height plotted along the line shown in (A).}
\end{figure}

  As a specific example of nano-vesicle, we consider an exosome.
  We quote from Ref.~\cite{pegtel2019exosomes}
  `` Exosomes are small, single-membrane, secreted organelles of
  $\sim 30$ to $\sim 200$ nm in diameter that have the same topology as the cell and are
  enriched in selected proteins, lipids, nucleic acids, and glycoconjugates. ''
  The exosomes that we consider here were collected from immortalized cell line and extracted
  following the procedures as described in Ref.~\cite{cavallaro2019label}.
  To measure the force-distance curve, it is necessary to fix an exosome on a
  transparent coverslip. This was done by  electrostatic coupling  to a PLL coated
  surface by incubating them at room temperature for one
  hour, see Ref~\cite{cavallaro2019label}.
  As an illustration, in  \fig{fig:exo}, we show a typical experimental
  measurement of the height above a flat surface as measured by the AFM.
  After being stuck to the flat surface the free surface forms a spherical cap. 
  To reproduce such experiments as closely as possibly we need to fix the
  vesicle to a flat surface.
  This is implemented by the Lennard-Jones potential:
  \begin{equation}
    \VLJ(r) \equiv 4\epsw\left[\left(\frac{\sigw}{r}\right)^{12}
      -\left(\frac{\sigw}{r}\right)^{6}\right]
    \label{eq:VLJ}
  \end{equation}
  What fraction of the vesicle is fixed to the flat surface is parametrized by the
  angle $\Theta_0$  (see ~\subfig{fig:exo}{B}) which is a parameter in our code.
  We choose a system of coordinates with its origin at the center of the vesicle
  and the $z$ axis pointing radially outward through the north pole.
  All the grid points on the surface whose polar angle is greater than $\Theta_0$
  are selected such that the sticking potential acts only on them, see \subfig{fig:exo}{B}. 
    %------------------------------------
\subsection{AFM tip}
To model the interaction between AFM tip and the vesicle,
We model the shape of tip as paraboloid and we use only the repelling part of the
Lennard-Jones potential:
\begin{equation}
      \VRLJ(r) \equiv 4\epsA\left(\frac{\sigA}{r}\right)^{12}
    \label{eq:VRLJ}
  \end{equation}
For every point on the vesicle we calculate the shortest distance of this point
to the paraboloid and use this distance as the argument of function $\VRLJ$ in
\eq{eq:VRLJ}.
\section{Dependencies}
The code requires the following:
\begin{itemize}
\item A c++ compiler. We have tested the code with gnu g++ version 11.2.0 on  x86\_64 CPU.
\item Hdf5 libraries for reading and writing data.
\item Python version 3.8 with scipy, numpy, h5py and numpy-quaternion installed.
\item For three-dimensional visualization we use VisIt~\cite{HPV:VisIt}.
\end{itemize}

\section{Typical workflow and test}
We have tested our code in LINUX operating system. We expect it
to work without any problem in any similar environment.
It may also work with WINDOWS although we have not tested this aspect.

\begin{figure*}
  \centering
  \includegraphics[width=0.90\textwidth]{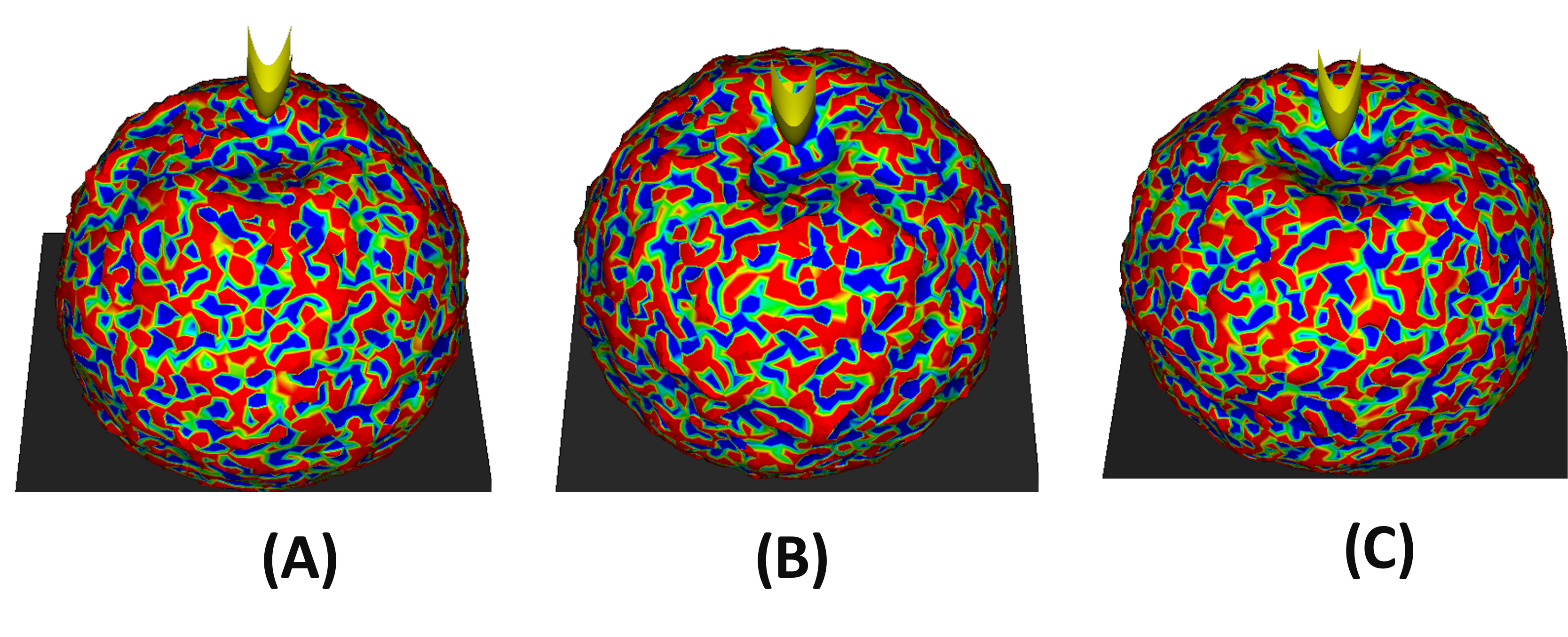}
  \caption{Representative snapshots from our code for three  different
    positions ($t_z$) of the AFM tip. The origin of our coordinate system  is at the
    center of the undeformed sphere and the radius of the
    undeformed sphere is unity. The colormap shows the signed 
  curvature~\eq{eq:laplace_beltrami}; red(positive) and  blue (negative).  
  (A) $t_z$ = 1.05, (B) $t_z$ = 0.9, (C) $t_z$ = 0.75.}
  \label{fig:wflow}
\end{figure*}

The github repository~\cite{memc} contains a a file named {\tt README.md}
that contains instructions to install and run the code.
In \fig{fig:wflow} we show three typical snapshots from our code
for three different positions of the AFM tip. 

In the github repository, we also provide a subdirectory called {\tt Examples}.
By executing the shell script {\tt execute.sh} in that directory
the user can run the code (without the AFM tip and the bottom plate). 
It takes almost $30$ minutes on Intel(R) Core(TM) i5-8265U CPU.
The run produces a probability distribution function (PDF) of the total energy
after $50,000$ number of Monte -carlo steps.
By running {\tt gnuplot plot.gnu} (this requires the software gnuplot)
the user can compare the PDF obtained by their run with
the PDF that we provide.
%--------------------------
%------------------------
% \begin{figure*}
%   \centering
%   \includegraphics[width=0.30\textwidth]{1p05_2_cr.png}
%   \qquad
%   \includegraphics[width=0.30\textwidth]{0p9_2_cr.png}
%   \qquad
%   \includegraphics[width=0.30\textwidth]{0p75_2_cr.png}
%   \caption{Representative snapshots from our code for three  different
%     positions ($t_z$) of the AFM tip. The origin of our coordinate system  is at the
%     center of the undeformed sphere and the radius of the
%     undeformed sphere is unity. The colormap shows the signed 
% 	curvature~\eq{eq:laplace_beltrami}; red(positive) and  blue (negative).  
% 	(A) $t_z$ = 1.05, (B) $t_z$ = 0.9, (C) $t_z$ = 0.75.}
%   \label{fig:wflow}
% \end{figure*}
% %------------------------
\acknowledgements
We acknowledge the support of the Swedish Research Council
  Grant No. 638-2013-9243 and 2016-05225.
The simulations were performed on resources provided by
the Swedish National Infrastructure for Computing (SNIC) at PDC center for
high performance computing.
%\bibliographystyle{apsrev4-1} 
% \bibliography{turb_ref}

%\appendix
\end{document}